\documentstyle[aaspp4]{article}
 
\lefthead{Elmegreen}
\righthead{Interacting Galaxies NGC~2207 and IC~2163}
\slugcomment{{\it Astronomical Journal}, in press August 2000}
\begin{document}
 
\title{HST Observations of the Interacting Galaxies NGC~2207 and
IC~2163\footnote{ Based on observations with the NASA/ESA Hubble Space
Telescope, obtained at the Space Telescope Science Institute, which is
operated by the Association of Universities for Research in Astronomy,
Inc. under NASA contract NAS5-26555 }}

\author{Bruce G. Elmegreen\altaffilmark{1},
Michele Kaufman \altaffilmark{2},
Curtis Struck\altaffilmark{3},
Debra Meloy Elmegreen\altaffilmark{4},
Elias Brinks\altaffilmark{5},
Magnus Thomasson\altaffilmark{6},
Mario Klari\'{c}\altaffilmark{7},
Zolt Levay\altaffilmark{8},
Jayanne English\altaffilmark{8},
L. M. Frattare\altaffilmark{8},
Howard E. Bond\altaffilmark{8},
C.A. Christian\altaffilmark{8},
F. Hamilton\altaffilmark{8},
K. Noll\altaffilmark{8}}

\altaffiltext{1}{IBM Research Division, T.J. Watson Research Center, P.O.
Box 218, Yorktown Heights, NY 10598; e--mail: bge@watson.ibm.com}
\altaffiltext{2}{Department of Physics and Department of Astronomy,
Ohio State University, 174 W. 18th Avenue, Columbus, OH 43210;
e--mail: rallis@mps.ohio--state.edu}
\altaffiltext{3}{Department of Physics and Astronomy, Iowa State University,
Ames, IA 50010; e--mail: curt@iastate.edu}
\altaffiltext{4}{Department of Physics and Astronomy, Vassar College, Poughkeepsie, NY 12604;
e--mail: elmegreen@vassar.edu}
\altaffiltext{5}{Depto. de Astronom\'{\i}a, Universidad de Guanajuato,
Apdo. Postal 144, Guanajuato, Gto. 36000, M\'exico; e--mail:
ebrinks@astro.ugto.MX}
\altaffiltext{6}{Onsala Space Observatory, S--439 92 Onsala, Sweden;
e--mail: magnus@oso.chalmers.SE}
\altaffiltext{7}{Columbia, SC 29206;
e--mail: mariok@cris.com}
\altaffiltext{8}{Space Telescope Science Institute, 
3700 San Martin Drive, Baltimore, MD 21218}

\begin{abstract} {\it Hubble Space Telescope} images of the galaxies
NGC 2207 and IC 2163 show star formation and dust structures in a
system that has experienced a recent grazing encounter. Tidal forces
from NGC 2207 compressed and elongated the disk of IC 2163, forming an
oval ridge of star formation along a caustic where the perturbed gas
rebounded after its inward excursion. Gas flowing away from this ridge
has a peculiar structure characterized by thin parallel dust filaments
transverse to the direction of motion.  The filaments get thicker and
longer as the gas approaches the tidal arm.  Star formation that occurs
in the filaments consistently lags behind, as if the exponential disk
pressure gradient pushes outward on the gas but not the young stars.
Numerical models suggest that the filaments come from flocculent spiral
arms that were present before the interaction.  The arms stretch out
into parallel filaments as the tidal tail forms.  A dust lane at the
outer edge of the tidal tail is a shock front where the flow abruptly
changes direction. Gas at small to intermediate radii along this edge
flows back toward the galaxy, while elsewhere in the tidal arm, the gas
flows outward.

A spiral arm of NGC 2207 that is backlit by IC 2163 is seen with {\it HST}
to contain several parallel, knotty filaments spanning the full width of
the arm.  These filaments are probably shock fronts in a density wave.
The parallel structure suggests that the shocks occur in several places
throughout the arm, or that the interarm gas is composed of spiral-like
wisps that merge together in the arms. Blue clusters of star formation
inside the clumps of these dust lanes show density wave triggering in
unprecedented detail. The star-formation process seems to be one of
local gravitational collapse, rather than cloud collisions.

Spiral arms inside the oval of IC 2163 have a familiar geometry
reminiscent of a bar, although there is no obvious stellar bar. The
shape and orientation of these arms suggest they could be the result of
ILR-related orbits in the $\cos(2\theta)$ tidal potential that formed
the oval. Their presence suggests that tidal forces alone may initiate a
temporary nuclear gas flow and eventual starburst without first forming
a stellar bar.

Several emission structures resembling jets or conical flows that are
$100-1000$ pc long appear in these galaxies. In the western arm of NGC
2207 there is a dense dark cloud with a conical shape 400 pc long and
a bright compact cluster at the tip, and there is a conical emission
nebula of the same length that points away from the cluster in the
other direction. This region also coincides with a non-thermal radio
continuum source that is $\sim1000$ times the luminosity of Cas A at
$\lambda=20$ cm.  Surrounding clusters in arc-like patterns may have
been triggered by enormous explosions.

\end{abstract}

Subject Headings: ISM: extinction -- 
ISM: jets -- 
ISM: dynamics -- 
galaxies: interactions --
galaxies: ISM
 
\section{Introduction}
\label{sect:intro}

The spiral galaxies IC 2163 and NGC 2207 are currently involved in a
near-grazing encounter (Elmegreen, et al. 1995a,b; hereafter Papers
III and IV).  Tidal forces distorted IC 2163 in the inplane direction,
forming a tidal arm with two velocity components on the anticompanion
side, and forming an intrinsically oval disk with an eye--shaped
(``ocular'') morphology.  Streaming motions in the oval are in excess of
100 km s$^{-1}$.  Tidal forces also distorted NGC 2207 perpendicular to
the plane, forming a strong, twisting warp.  The H~I in both galaxies has
a large velocity dispersion of 30 - 50 km s$^{-1}$ and is concentrated in
unusually large clouds ($10^8$ M$_\odot$; Elmegreen, Kaufman \& Thomasson
1993; hereafter Paper II).  Numerical simulations reproduced these
peculiar structures and internal velocities with a prograde encounter
affecting IC 2163 and a perpendicular encounter affecting NGC 2207 (Paper
IV). The closest approach was around 40 My ago at a separation of only
$\sim2$ radii.  This makes IC~2163 and NGC~2207 ideal for studying close
encounters between galaxies and how they might trigger bars or starbursts.

Three systems of interacting galaxies (IC 2163, NGC 2535, and NGC 5394)
have now been studied by our group to see how structures resulting from
prograde, inplane, nearly-grazing encounters evolve in time (Papers
III, IV; Kaufman et al. 1997; Kaufman et al. 1999). IC 2163 is the
interaction which is least evolved. In all three cases, the prograde
galaxy has two long tidal arms with a large arm/interarm contrast,
misalignment between the kinematic and photometric axes indicative of
an intrinsically oval disk, and widespread high velocity dispersions in
the H~I gas.  NGC 2535 is an ocular galaxy but, unlike IC 2163, it has
no enhanced HI or radio continuum emission on the rim of the oval; this
indicates it is in a later phase than IC 2163 (Elmegreen et al. 1991;
hereafter Paper I). There is also evidence for mass transfer from NGC
2535 to its small starburst companion. NGC 5394 has no ocular structure,
but bright spiral arms in the inner disk and a nuclear starburst, making
it more evolved than NGC 2535.

To study these grazing encounters in more detail, we observed the IC
2163/N2207 pair with the WFPC2 camera on the Hubble Space Telescope
(HST) in May 1996 and November 1998. Based on these observations, this
paper describes (a) an improved understanding of the formation mechanism
of the tidal tail in IC 2163, with a detailed numerical simulation of
the gas (Sect.  \ref{sect:tail} and \ref{sect:tailH}); (b) a new view
of the density-wave structure and star formation in a spiral arm, as
seen in a backlit portion of NGC 2207 (Sect. \ref{sect:foreground}; (c)
bright star formation associated with a strong nonthermal radio source
on an outer spiral arm (Sect. \ref{sect:rcsource}); (d) the possibility
that interactions can induce transient nuclear gas inflow via bar-like
hydrodynamics, even if a genuine bar is absent (Sect. \ref{sect:nucsp});
(e) peculiar jet--like or curved emission features (Sect. \ref{sect:pec});
and (f) the HI -- star formation connection (Sect. \ref{sect:hi}).
Other studies of the star--forming regions and dust opacities will
be reported elsewhere (D. Elmegreen et al. 2000, hereafter Paper V;
see also Berlind, et al. 1997).  The HST observations also revealed
peculiar dust spirals in the nucleus of NGC 2207; these were analyzed
in terms of acoustic instabilities by Elmegreen et al. (1998).

\section{Observations and Data Reduction}

The galaxies IC 2163 and NGC 2207 were observed in 9 orbits with the
{\it Hubble Space Telescope} WFPC2 camera using filters F336W (U band),
F439W (B band), F555W (V band), and F814W (I band). Four exposures in each
band concentrated on a field in the northwest, and four more exposures
of each of two fields were made for the central and eastern parts. The
images were dithered between each exposure, and the exposure times were
500 s in U, 500 s in B, 160 s in V, and 180 s in I. The average scale
is $0.0995^{\prime\prime}$ per pixel for the wide field images and
$0.0455^{\prime\prime}$ per pixel for the planetary camera images.

From these 48 individual pipeline-processed exposures (four filters,
three pointings, two dither positions per pointing, and two CR-splits
per position), cosmic ray split pairs were combined for each dither
position chip by chip. Individual WFPC2 chips were then combined for
each dither position into $2907\times1486$-pixel mosaics. The dither
positions were registered and combined, resulting in a single image
for each filter and pointing (twelve images). A 3-color composite image
was produced (using IDL) for each pointing and the three pointings were
registered and mosaicked (using Adobe PhotoShop) into a single image.
The initial step in creating each 3-color image involved clipping
and logarithmically scaling the intensities in each filter.  Then each
filter was assigned, in chromatic order, to a color channel: red=log(I),
green=log(V), blue=log(B+U).  The 3 channels of the mosaicked image were
combined using the ``screen'' algorithm in PhotoShop. The channels were
adjusted in order to match more closely the visual color of the central
wavelength of each filter. This produced an overall bright image with
a neutral color background, and it increased the intensity and color
dynamic ranges in order to emphasize individual features.  The noisier
PC chip was Gaussian smoothed to match the texture of the other chips.
Finally, the remaining artifacts (hot pixels, chip seams, etc.)
were removed using Photoshop.

For comparison with H~I and radio continuum images, a B--band mosaic
was transformed to right ascension and declination coordinates. The
uncertainty in the registration to absolute coordinates is about $\pm
0.8^{\prime\prime}$, partly as a result of uncertainties in the positions
of the Guide Stars and secondary standard stars and partly as a result
of the standard deviations of the plate solution. The B-band mosaic in
absolute coordinates was rotated 10.7$^\circ$ counterclockwise relative
to the color mosaic displayed in Figure 1 below.

\section{Basic Morphology of the Interacting Galaxies}
\label{sect:basic}

A mosaic of three {\it HST} WFPC2 pointings of the galaxy pair
IC~2163/NGC~2207 is shown in Figure \ref{fig:mainimage}, with North up
relative to the figure caption. A digitized version of this figure may
be found at the Hubble Heritage web site (http://heritage.stsci.edu).
Here we comment on several unusual features. The relevant parameters
for these galaxies are listed in Table 1.  At the distance of 35 Mpc
(H=75 km s$^{-1}$ Mpc$^{-1}$), $1^{\prime\prime}$ corresponds to 170 pc.

The morphology, interaction dynamics, and orbital history of this system
were discussed, along with numerical simulations of each galaxy, in Papers
III and IV.  These previous simulations considered stars only, without
gas. This was enough to get the basic disk structure. A new simulation
of the tidal arm discussed below includes gas.  The orbit that we fit
previously had the smaller galaxy, IC 2163 (presently in the east), pass
from the near side of NGC 2207 to the far side several hundred million
years ago, at a point in the west about two NGC 2207-radii away from the
center of NGC 2207. IC 2163 then moved behind NGC 2207 in an easterly
direction, passing perigalacticon at a distance of about two of its own
radii some $40$ My ago. IC 2163 is currently moving southeasterly. These
relative distances are uncertain because they depend on the sizes and
masses of the halos around each galaxy. The orbit is constrained by the
internal structures and velocities of the galaxies, by their similar
line-of-sight velocities (which implies that they are presently moving
nearly parallel to each other in the sky plane), and by the extended
pool of optical and H~I emission that lies south of NGC 2207.  This pool
seems to be the point of the previous disk crossing, where IC 2163 passed
through the outer disk plane of NGC 2207, now rotated clockwise by about
1/4 revolution.

During this interaction, IC 2163 experienced the tidal force from NGC 2207
in a prograde, inplane direction, and subsequently developed an ocular,
or oval-like, caustic structure midway out in the disk. Such ocular
structure is the result of a rapid inward motion of both stars and gas
in the disk of IC 2163, initiated by the stretching and compressing
action of the tidal force and amplified by self-gravity in the disk
of IC 2163 (Sundin 1989; Paper I; Donner, Engstrom \& Sundelius 1991).
The inward motion causes an azimuthal speed-up in the counter-clockwise
direction, and the radial motion then bounces because of angular momentum
conservation. The oval marks the inner extent of this radial excursion.
The minor kinematic axis of IC 2163 lies $\sim117^\circ$ away from the
minor photometric axis, indicating an intrinsically oval disk.

There is a peculiar streaming motion along the oval of about 65 km
s$^{-1}$ counter-clockwise (Paper III).  The line--of--sight component
of the peculiar velocity increases to $\sim150$ km s$^{-1}$ as the
H~I gas and stars enter the tidal arm from the northwest, and then it
decreases suddenly to $<50$ km s$^{-1}$ toward the outer edge of the
tidal arm (Fig. 10, Paper III).  This motion was evident in the channel
maps of Paper III (Fig. 4) and in a velocity-position map (Fig 9),
but in the line-integrated velocity map the streaming motion was less
obvious. Nevertheless, in the velocity field image in Paper III (Fig. 11),
the highest velocity gas occurs at the eastern end of the northern eyelid,
displaced $63^\circ$ in position angle from the kinematic minor axis
rather than $90^\circ$ for a normal rotation curve.

In Paper III we referred to the eastern tip of the oval, where the
non-circular motion peaks, as the launch point for gas leaving the
caustic and entering the tidal arm. The gas streams outward here mainly
because it is traveling too fast for a circular orbit, but tidal forces
also produce some outward acceleration.  By the time the gas reaches
the southern edge of the tidal arm, its peculiar motion has slowed down
because of the gravity of the galaxy and large scale shocks.

During the formation of the tidal tail in IC 2163, NGC 2207 experienced a
perpendicular tidal forcing from the gravity of IC 2163. This perturbation
seems to have created an inward-propagating tidal warp with an overall
spiral shape, as determined by shear and disk self-gravity (Paper IV). The
corresponding warp in the velocity field was clearly seen (Paper III),
although the optical disk shows little evidence for it. The warp is such
that the side of NGC 2207 closest to IC 2163 in projection is warped
toward it by about 9 kpc compared to the average inner disk plane.

\section{Peculiar Dust Structures}
\label{sect:dust}

Dust features are sensitive indicators of shock fronts because even small
compressions can change the gas from optically thin to optically thick.
There are many interesting dust structures in these galaxies as a result
of the complex and unusual dynamics of the interstellar media.  The high
resolution and the backlighting of the outer spiral arm of NGC 2207 by
IC 2163 provide a unique opportunity to study these structures.

Two peculiar dusty regions will be highlighted here: the parallel
striations in the tidal tail of IC 2163, and the foreground spiral arms
in NGC 2207.  A dense dark cloud in the far western arm of NGC 2207 will
be considered in the next section.

\subsection{Parallel Dust Filaments in the Tidal Tail of IC 2163}
\label{sect:tail}

The tidal tail in IC 2163 begins at the eastern end of the inner oval
with a broad and shallow distribution of starlight.  Gas and stars flow
outward and downward, ending abruptly in the southeast along a curving
arc where there is a dust lane, and extending northwest along the length
of the arm for $\sim15$ kpc. A dwarf galaxy with unknown velocity lies
just off the tip of the tidal arm (the presence of globular clusters
around this nucleated dwarf elliptical suggests it is not related to the
interaction -- see Paper V).  Along the tidal outflow, there are numerous,
nearly-parallel dust filaments perpendicular to the flow direction
$\sim0.5^{\prime\prime}=85$ pc thick, spaced by about $2^{\prime\prime}$.
The filaments are darker than the surrounding regions by 0.2--0.8 mag
in V band, suggesting that the midplane densities are enhanced by a
factor of $\sim5$ compared to the inter-filament gas (Paper V). The
filament widths increase toward the outer edge of the tidal arm, where
they take the form of two long, dense dust lanes, still parallel to each
other. There is a slight tendency for the highest opacity dust streamers
to occur at smaller radii (Paper V).

In our previous simulations considering only stars (Paper IV), the
broad plateau of light on the inside part of the eastern tidal arm is a
region of outward streaming stars. The outward motion is 150 km s$^{-1}$
in places and is the result of angular momentum conservation correcting
for the inward plunge that these same stars recently made in response
to the tidal force and self-gravity (Papers I and IV). The stars stream
outward and form the tidal arm, where they slow down to a lower streaming
speed of 50 km s$^{-1}$.  These model velocities for stars were chosen
to match the observed velocities for the gas. Ground-based observations
could barely resolve the dense, double dust lane at the edge of the arm,
and they could not see the fainter filaments in the broad plateau.

Paper III briefly considered the possibility that the split in the
tidal arm of IC 2163 was the result of dust, but we favored a different
interpretation in which it resulted from two separate stellar arms
created for a short time during the interaction. One of these arms was
a caustic, reproduced by the model, and the other was a normal tidal
arm. Now we see with higher resolution that the split is from dust
and conclude either that there was no caustic arm or that the present
epoch is not the correct one for viewing such a short-lived feature.
The two dust lanes are hydrodynamic features, probably shock fronts,
and could not be modeled before with our pure-stellar code. Thus we
performed a new model with gas.

This new model is for a nearly-inplane, prograde interaction (25$^\circ$
inclination) between a gaseous disk galaxy and a point source. The tidal
arm part of the model is shown in Figure \ref{fig:model}. The ocular
structure (i.e., the caustic oval in the inner disk) is produced again
because this is a general property of prograde nearly-inplane encounters.
Here we highlight the parallel filaments in the outward-streaming regions
of the tidal arm, which resemble the filamentary dust structures in Figure
\ref{fig:mainimage}. A more detailed model of the whole two-galaxy system,
including a fuller analysis of the gas motions between the galaxies and
the warp in NGC 2207, will be deferred to a later paper.  The filamentary
arms in the tidal tail of IC 2163 are present in the detailed study too.

In the present model, the gravitational potential of IC 2163 is dominated
by a dark matter halo that is effectively rigid over the relatively short
duration of this interaction. Smoothed particle hydrodynamics (Struck
1997; Kaufman et al. 1999) with 18,000 particles calculates the dynamics
in the disk, which is assumed to be pure gas, using an adiabatic equation
of state with heating and cooling. The initial disk was in rotational
equilibrium with small thermal motions. It had an exponential density
distribution over four scale lengths, and a thickness perpendicular
to the plane equal to one scale length. The thermal terms lead to the
formation of a multiphase initial gas. The companion is assumed to be a
pure halo with half the scale length of the IC 2163 halo and 1.25 times
the total IC 2163 mass. The observed ratio of H and K band luminosities
is 1.6 (Kaufman et al. 1997).

Figure \ref{fig:model} shows three face-on views of the model disk and a
projected view: (a: top left) is the map of model points at the beginning
of the interaction, using color to represent initial variations in the
particle positions with galactocentric radius; (b: top right) is the point
map at a time representative of the current NGC 2207/IC 2163 system, with
the same color scheme; (c: bottom left) shows the velocity vectors at the
same time as in (b), and (d: bottom right) shows the density at the same
time again, but rendered from a 160x160 array with color proportional to
the density of the model points inside each pixel (blue is low density
and red is high density), and projected by an inclination of $30^\circ$
to resemble IC~2163 in the sky. The center of the companion, NGC 2207,
is off the figure in b, c, and d.

The tidal tail in the model has the same shape as the observed tidal tail,
and an intricate network of parallel filaments is in the broad part of
this tail (seen best in the lower right panel).  These filaments are
stretched flocculent spiral arms that were present in the disk before
perigalacticon (cf.  Figure \ref{fig:model}a). The model generates only
flocculent spirals because of the relatively high mass of the halo and the
low temperature of the pure-gas disk (see Elmegreen \& Thomasson 1993).

The velocity vectors in Figure \ref{fig:model}c indicate there is a
shock front along the outer edge of the tidal arm, where the velocities
abruptly change. This front also corresponds to a region of enhanced
density, and it is located at the same place as the dense dust lane
in the tidal arm of IC 2163.  This result indicates that the dust lane
running along the edge of the IC 2163 tidal arm is a shock front where
the velocities change from outward to inward streaming.  Not all of the
tidal arm material is unbound from the galaxy: the inner portion returns
to orbit the main disk after it shocks in the dust lane.

In Figure \ref{fig:mainimage}, many of the filaments in the tidal arm of
IC 2163 have bright star clusters adjacent to them. These clusters are
typically on the inside edges, toward smaller galactocentric radii. This
is true for the small and faint dust filaments, as well as the dense dust
lanes near the edge. The systematic displacements of the star clusters
relative to the tidal arm dust filaments is an important check on the
dynamics of the gas. Presumably these clusters formed in the dust lanes
and then became ballistic after they formed. The dust and gas are not
ballistic, however, and respond to the pressure forces. Thus the relative
displacement of the dust and the clusters indicates the direction of
the pressure gradient. To get the dust filaments systematically outside
the clusters, the gas pressure must be decreasing with increasing
galactocentric radius, as expected for an exponential disk.

The displacement between stars and dust in this interpretation is unlike
the situation in an idealized spiral arm shock. There the displacement
inside corotation (where the stars are systematically outside the dust
lanes) is the result of a spiral wave and associated shock front (the dust
lanes) that move slower in the azimuthal direction than {\it both} the gas
and the stars. The dust lane in a normal spiral is newly shocked material
that did not yet form stars, while the stars and the gas clouds in which
they formed are both ballistic particles that move ``downstream.'' The
dust lane material has just been shocked, and so is moving at its most
negative radial speed.  It is also at its largest galactocentric radius,
having just streamed outward through the interarm region. As the dust lane
gas moves inward and along the arm, it presumably forms clouds and stars
that feel a Coriolis force deflecting them outward. In this way, their
inward motion is converted into an azimuthal motion, and the young stars
begin to emerge from the region of the shock toward the interarm region.
For this normal spiral arm, the stars and the dust lane are displaced
from each other because these two features are at different parts of
their epicycles: the dust lanes are where the radial motion is greatest
inward and the azimuthal motion relatively small, while the stars and
associated clouds just downstream from the dust lanes are entering the
parts of their epicycles where the azimuthal motion is greatest in the
prograde sense.  This juxtaposition creates the illusion that the stars
formed out of gas that is located in the dust lane right next to them,
but this is not the case. Instead, the young stars formed in gas clouds
that are moving downstream with them, and the dust lane next to these
stars is newly shocked material that has just arrived in the wave.
The process of star formation in spiral arms will be discussed further
in Section \ref{sect:foreground}.

For the dust lanes in the tidal tail of IC 2163, the situation should be
different. This is a stretched and distorted part of the former galaxy,
and the motions through the arm are largely radial. In this case, the
radial disk gradient of the interstellar pressure contributes to the
force on the gas, pushing it faster than the stars which recently formed.

\subsection{Comparison Between the Dust and Atomic Hydrogen in the Tidal
Arm of IC 2163}
\label{sect:tailH}

Papers III and IV found H~I streaming motions in the tidal arm of IC
2163 that are consistent with the outward flow of gas expected from the
arm formation models. Figure \ref{fig:himap2163} shows the H~I overlaid
on the B band {\it HST} image. The triangles correspond to the ridge
of maximum streaming speed, where the line-of-sight velocity is fairly
constant at 2970--2990 km s$^{-1}$ (the galaxy systemic velocity is
$2765\pm20$ km s$^{-1}$). The pentagons correspond to a ridge of lower
speed, ranging from 2780 to 2930 km s$^{-1}$ (Paper III, Sect. 4.2). The
plus symbols are along the dividing line between the high velocity and
the low velocity H~I ridges, where the H~I intensity is low.

In Paper III, we expected the dividing line to lie at a dust lane, where
the high speed stream suddenly shocked into the low speed gas and the H~I
got converted into H$_2$. This does not appear to be the case. Instead,
the streaming ridge corresponds to a diffuse, optically faint region
on the northern part of the tidal arm, and the low-speed gas in the
south corresponds to the unresolved combination of the two main dust
filaments. The dividing line (plus symbols in Fig. \ref{fig:himap2163}) is
unrelated to the main dust lanes because it has a different curvature. The
H~I arm also has an S-shaped wiggle at $\alpha_{1950}=6^h14^m25^s$ --
$6^h14^m27^s$, where the high and low velocity components of the arm
appear to merge.  The dust lanes do not show this structure.

The new model in Figure \ref{fig:model} explains the origin of these
moving streams. Most of the high-velocity stream in the northern part
of the tidal arm is from the original outer disk of IC 2163, which was
H~I-dominated and had a low stellar density before the encounter. The
optical faintness of the northern part of the arm is consistent with an
outer disk origin. The H~I gas in this region moves quickly on its way to
the tip of the arm where it may form a large gas pool (cf. Paper II).
The intermediate zone of low H~I emission (plus symbols) is mostly
stellar, and it comes from the intermediate- to outer-radii of the
optical disk of the former IC 2163. This part of the tidal tail moves
in a southeasterly direction through the arm at a steady projected
speed until the gas shocks along the curved arm edge. Then it forms the
prominent parallel dust lanes at a relatively low line-of-sight velocity,
corresponding to some of the gas moving inward.

The asymmetry in the fall-off of the H~I intensity on the northern
and southern sides of the tidal arm is consistent with this picture:
the northern H~I is extended gas that may have come all the way from
the companion side of the IC 2163 outer disk, whereas the southern H~I
is mostly shocked gas at the leading front of the tidal arm. Figure 3b
of the model in Paper II illustrates this morphology.

\subsection{Foreground Dust in the NGC 2207 Spiral Arms}
\label{sect:foreground}

The background lighting by IC 2163 provides a unique view of the dust
lane and star formation in an outer spiral arm of NGC 2207. What appears
in a ground-based image to be a single dust lane in a spiral arm of NGC
2207 is seen at higher resolution to consist of 4 to 7 nearly parallel
dust streamers that span the full width of the arm. The V-band magnitude
decrements in many of the dust features of these arms are in the range
from 0.4 to 1.5 mag (Paper V).

The spiral arm dust lanes in the {\it HST} image of NGC 2207 show an
intricate internal structure reminiscent of Galactic cirrus clouds (Low
et al. 1984) or other diffuse interstellar structures in the Milky Way,
but on a much larger scale. There are also blue stars or clusters adjacent
to many of the dust clouds, as if they just formed there. These spiral
arms are density waves in NGC 2207, and their dust structures presumably
delineate the shocks in these arms, as in the standard theory (Roberts
1969). However, the shocks are not smooth, and they are not just clumpy
in the usual sense either (e.g. Combes \& Gerin 1985; Roberts \& Steward
1987; Elmegreen 1988). They are mostly composed of long, knotty filaments,
which in some places run side by side in parallel streaks. Such structure
suggests that the density wave shock occurred in several separate places
along the width of the arm, or that the interarm gas is not in the form
of spherical clouds, but filamentary like in M51 (Block et al. 1997).

We do not see the old stars in these arms because they are too faint
compared to the background disk of IC 2163. We can see them in other
parts of the foreground arms, however, such as the region north of IC
2163 before the arms cross in front of the disk. There are many faint red
stars in these parts too, alongside young clusters and other dust lanes.

The blue stars in the foreground spiral arms are interesting because of
what they tell us about the processes by which density waves trigger or
organize star formation. This is the first example in an external galaxy
where we can see such triggered star formation in projection against a
background source. We see at this resolution (1 pixel $=$ 17 pc) blue
stellar-like images mixed with the dust. These blue objects are probably
young clusters (Paper V).

The star formation process looks normal at this perspective. The filaments
contain clumps, and the clumps form stars. This is the same morphology
as in local regions that are much smaller, such as Taurus. We cannot
see any dust structure that might be indicative of cloud disruption,
such as comet tails pointing away from the blue clusters, but the dust
opacity may be too low for such detail on these scales.

The tidal forces acting on NGC 2207 are primarily perpendicular to its
disk, and this galaxy is not flocculent, so it is unlikely that the
parallel dust streamers in the spiral arm are from tidally stretched
flocculent arms, as in IC 2163. The parallel dust filaments in NGC 2207
look similar to parallel dust filaments in the arms of M81. These M81
filaments are often at the upstream and downstream edges of the strings of
giant HII regions in the arms, and may be parts of dense shells (Kaufman,
Elmegreen, \& Bash 1989), but it is also possible that M81 and other
grand-design spirals have the same multi-stranded shock structures in
their arms as NGC 2207.

Not all of the spiral arms in NGC 2207 have parallel dust structure.
In the west, midway out in the optical disk, the dust lanes cut through
the arm like spurs, but farther to the west, in the outer arm, some of
the dust lanes are parallel again.  Perhaps this change to spurs marks
an orbital resonance.

\section{The Dense Dark Cloud and Star-Forming Region in the Far West
of NGC 2207}
\label{sect:rcsource}

There is a peculiar dust region that appears to be associated with star
formation in the far western part of the outer spiral arm of NGC 2207.
It is the site of an intense radio continuum source, as shown in Figure
\ref{fig:rcsources}, and the most luminous H$\alpha$ source in the
system (see map in Paper V).  An enlargement of the {\it HST} image is
labeled feature i in Figure \ref{fig:jets}. In Paper III, we conjectured
that this radio continuum source was a background radio galaxy, not
associated with NGC 2207, because if it were at the distance of 35 Mpc
its luminosity would be unlike anything else in either galaxy. Now,
its coincidence with a dense dust cloud and bright young star cluster,
and the similarity between the velocities of the associated H$\alpha$
emission and the H~I emission from NGC 2207 in this vicinity, make the
association of this radio continuum source with NGC 2207 more likely.

With a resolution of $1.0^{\prime\prime}\times 1.9^{\prime\prime}$
FWHM, Vila et al. (1990) detected a compact, radio continuum core with
a deconvolved Gaussian size of $\sim 1^{\prime\prime}$ (marked with a
plus sign in Fig. 4) at the location of the star cluster, surrounded by
a more extended radio component. The core has flux densities of $S(20)$
=3.4 mJy at $\lambda$ 20 cm and $S(6)=1.4$ mJy at $\lambda$ 6 cm and a
spectral index $\alpha$ of --0.7. Integrating over a $7.5^{\prime\prime}$
= 1.3 kpc region, they measured $S(20)=10.3$ mJy, $S(6) = 3.4$ mJy and
$\alpha = -0.9$. The lower resolution radio continuum observations in
Paper III found even more extended emission here, with $S(20)$ = 22 mJy
and a deconvolved size of $2.5 \times 2.2$ kpc (FWHM). Thus the core
has a radio continuum luminosity 300 times that of Cas A; the core plus
extended component forms a big, nonthermal radio source with a radio
continuum luminosity $\sim 1500$ times that of Cas A if the spectral
index of --0.9 applies throughout (taking the $\lambda$ 6 cm luminosity
of Cas A as $7 \times 10^{17}$ W Hz$^{-1}$ from Weiler, et al. 1989).
Although comparable in radio luminosity to the most luminous radio
supernovae (e.g. SN 1986J, SN 1988Z; Van Dyk, et al. 1993), the large
linear size of the source in NGC 2207 suggests it is something else.

The H$\alpha$ source in this region has a luminosity (from Paper V) of
$1.4 \times 10^{40}$ erg s$^{-1}$ and a diameter of $6^{\prime\prime}$.
Uncorrected for reddening, the H$\alpha$ flux is equivalent to S(6) =
0.13 mJy of optically thin free--free emission if $T_e = 10^4$ K. This
is much smaller than the measured $\lambda$ 6 cm flux density from
approximately the same region. Its H$\alpha$ luminosity uncorrected
for reddening is similar to that of 30 Dor (Kennicutt \& Hodge 1986)
and the brightest HII region in M51 (Van der Hulst et al. 1988). The
strong nonthermal emission from the NGC 2207 source suggests that some
of the H$\alpha$ flux may be from shocks or synchrotron sources rather
than photoionization.  A significant amount of H$\alpha$ could be occulted
by the dark cloud, but not all of it because we still see star clusters,
and the radio continuum and H$\alpha$ sources are larger than the cloud.

The structure of the region in the {\it HST} images is also peculiar
because of a V-shaped feature, possibly a conical outflow, that has a
bright star cluster at the apex and opens up to the north. This V-shape
seems to extend on its western side to a bright star-like object in the
NNW (which may not be related to it) at a distance of $\sim500$ pc from
the central cluster. A dense dust cloud trails off in the other direction.
If this V-shape comes from a conical outflow, then there may be a peculiar
and energetic star or collapsed object in this region, possibly with an
accretion disk and jet. An x-ray survey might reveal a compact source.

To the southeast of the dense cloud and central star cluster, there are
a number of smaller star clusters (Fig. \ref{fig:jets}i) that appear
to be distributed in concentric arcs $\sim400$ pc long.  A similar
pattern was found in an equally intense region of star formation in
the galaxy NGC 6946 (Elmegreen, Efremov, \& Larsen 2000), where several
cluster arcs are inside and at the edge of a cleared $\sim600$-pc cavity
surrounding a $\sim15$ My old globular cluster.  If these arcs are not
optical illusions (e.g. Bhavsar \& Ling 1988), then the clusters could
have been triggered in expanding partial shells.  The large scale of this
process, the unusual morphology, and the strong nonthermal radio emission
from the NGC 2207 source suggest peculiar conditions.  Considering the
arc sizes, the explosions that triggered the clusters would have been
extremely energetic, more than just single supernovae, and they were
possibly one-sided since the apparent cluster arcs are not complete
circles. It may be that a few extremely massive stars and their hypernovae
(Paczy\'nski 1998) are involved, in which case there could have been a
gamma-ray burst from these regions several tens of millions of years ago
(see also Efremov 2000). The one-sided nature of the cluster distribution
could also be the result of an initial gas concentration on that side.

There is no significant excess of H~I gas at the location of this dust
cloud and star-forming region (see Figure \ref{fig:himap2207} below). An
H~I cloud exists slightly to the south, along the spiral arm in a large
darkish region, and another H~I cloud is to the northeast, in the dark
interarm region. These two clouds could be pieces of the envelope of a
former giant molecular cloud centered on the dust feature. To the south
along the same arm, the next big star-forming region does have a prominent
H~I cloud associated with it. Thus the more intense star-forming region
associated with the dense dust cloud and the strong radio continuum
source could have blown apart the low density parts of its cloud. We
will discuss the other H~I structure in these galaxies in more detail
in Section \ref{sect:hi}.

The mass of the dust cloud can be estimated from its size ($\sim300$
pc $\times140$ pc) and apparent darkness. Paper V measured brightness
deficits of $\Delta m_U=1.78\pm0.3$, $\Delta m_B=1.73\pm0.3,$ $\Delta
m_V=1.24\pm0.3,$ and $\Delta m_I=1.38\pm0.3$ mag in the dark cloud. Since
these values are all about equal, the dark cloud is optically thick
in even the I band. If we assume an average visual extinction of 3 mag
(twice $\Delta m_V$ because of foreground stars), then the mass would
be $\sim10^6$ M$_\odot$.

\section{Inner Disk Spirals in IC 2163}
\label{sect:nucsp}

Paper III speculated that the nuclear region of IC 2163 might have a weak
bar, or possibly be forming a bar now as the result of the interaction,
as predicted by Noguchi (1987), and shown again by Gerin et al. (1990)
and Paper I. The {\it HST} image shows no evidence for such a bar but
there is a spiral arm system inside the oval that is elongated to make
the overall structure resemble the inner Lindblad resonance (ILR) region
of a bar.

The elongation of the two inner spirals arms is actually much more
pronounced in the image than it is in reality because the galaxy is
inclined with a line of nodes nearly parallel to the {\it minor} axis.
This is the opposite of what most galaxies have: usually the projection
line of nodes is along the major axis. For IC 2163, the kinematic minor
axis is approximately the same as the morphological major axis in the
oval region (Paper III). The kinematic minor axis from H~I velocities has
a position angle of $155^\circ$, and the morphological major axis from
optical and H~I isophotes has a position angle of $128^\circ$, which is
only $27^\circ$ different. The inclination is $30^\circ$ -- $40^\circ$
from the model fits. Thus the bright oval in IC 2163 is actually more
elongated than the image shows (cf. Fig. \ref{fig:model}), by a factor
of $1.2-1.3$, and the spirals inside the oval are more circular.

Figure \ref{fig:ai} displays the radial behavior of the arm/interarm
contrast of the inner--disk spirals in V and I-bands. The arm contrast
is nearly constant with radius, which is unusual for the inner parts of
density-wave arms (Elmegreen \& Elmegreen 1995).

There is a curious dynamical property of this interaction that may
point to the origin of the inner--disk spiral and even suggest a common
mechanism for close interactions to trigger nuclear gas accretion and
nuclear starbursts. This property follows from the relative position of
the companion, NGC 2207, and the rotation curve inside IC 2163.

Figure 13 in Paper I showed the rotation curve of IC 2163, from the
H~I data. It is steeply rising as a solid body between the center and
$\sim7^{\prime\prime}$, and then it is more slowly rising in a steady
fashion after that. This inner steep rise means that there will be an
inner ILR somewhere inside $7^{\prime\prime}$ and an outer ILR somewhere
outside $7^{\prime\prime}$ for any bi-symmetric spiral pattern that has
a sufficiently low pattern speed.  The semi-minor axis of the oval is
about $15^{\prime\prime}$ and the inner radius of the inner spiral is
about $5^{\prime\prime}$.  Thus the inner spiral could extend between
the inner and outer ILRs.

For a conventional bar potential with an ILR, there is a shock, or dust
lane, parallel to the bar on the leading side and then a twist to a ring
or inner spiral at around the ILR (Athanassoula 1992).  If there is both
an outer ILR and an inner ILR, then there will often be a 90$^\circ$
turn of this shock between the two resonances (Sanders \& Huntley
1976). The spiral inside the oval of IC 2163 turns about 90$^\circ$,
and may have its outer radius at the outer ILR, which would then be at
the inner extent of the oval. The spiral could have its inner radius
at the inner ILR.  The whole oval would then be showing the response
of the galaxy IC 2163 to a bar-like or $\cos(2\theta)$ potential: the
long axis of the oval, which has an axial ratio of about 3:1 corrected
for inclination, is along the ``bar'', where the inward radial force
and $\cos(2\theta)$ are maxima. The inner near-circular spiral, which
extends from the inner minor axis of the ``bar'' to a smaller radius,
twisting 90$^\circ$ along the way, would be the ILR shock feature. This
bar-like structure presumably gave IC 2163 the de Vaucouleurs type of
SB(rs)c pec, even though there is no bar in the classical sense.

An interesting thing about this system is that the tidal potential
from NGC 2207 has the correct orientation to contribute to the ``bar''
potential in IC 2163, along with the oval itself. The tidal force from
NGC 2207 pinches inward on IC 2163 in a direction that is perpendicular
to the line connecting the two galaxies, and it pulls outward on IC
2163 in a direction that is parallel to this intergalactic line. This
makes a $\cos2\theta$ forcing, much like in a bar (Combes 1988). For
a real bar, the perturbation in the inward direction, relative to
a circular potential, is along the bar, and the perturbation in
the outward direction, relative to a circular potential, is on the
minor axis of the bar. Thus the inward tidal pinch and outward tidal
pull from NGC 2207 makes IC 2163 feel like it has a bar oriented in a
direction perpendicular to the line connecting the two galaxies. This
is in fact the mean orientation of the oval during the last $\sim80$
My, which is the total time from the current position $\sim40$ My after
perigalacticon back to the symmetric position prior to perigalacticon.
Recall that IC 2163 is quickly moving along the backside of NGC 2207 in a
direction roughly from west to east and parallel to the disk of NGC 2207.
The transient deformation of the disk into the oval accentuates this
tidal force, increasing the bar-like potential, and this transient bar
can drive ILR spirals.

What makes the tidally-induced bar potential so prominent in IC 2163
is the fortunate circumstance of catching the galaxy pair close to
perigalacticon, and the extreme proximity of the encounter (perigalacticon
was estimated to be $\sim2.3$ times the radius of IC 2163 in Paper
IV). Presumably the bar-like response in IC 2163 will last only during
the closest approach, unless IC 2163 actually forms a permanent bar
during this time, which is possible (Paper I).

The corotation position of the bar-like potential in this interpretation
is the companion galaxy, NGC 2207. This is a large distance from IC
2163, and it gives a low pattern speed (approximately $5.6$ km s$^{-1}$
kpc$^{-1}$). For such a pattern speed, the inner and outer ILRs
should both exist for the rotation curve of IC 2163, and they should
be sufficiently well separated to account for the inner-disk spiral
structures.  Unfortunately, the H~I rotation curve is not sufficiently
accurate, considering the streaming motions, to determine these inner
and outer ILR positions any better.

A transient, tidally-induced, bar-like potential in an interacting galaxy
should be able to drive mass inflow to the nucleus in much the same
way as a permanent bar structure. The bar produces torques on the gas,
and the gas moves to the center, primarily along the dust lanes and ILR
spirals, with the resulting loss of angular momentum. If the encounter
is too weak and fast to make a bar (the criteria for bar-making were
discussed in Paper I), then the system will end up with a starburst in
the nucleus of one or both galaxies (if they both felt an inplane tidal
force), and there will be no evident bar structures at late times to
indicate how the gas got to the center so quickly. In this way, tidal
interactions may trigger gas accretion and nuclear starbursts in a
variety of seemingly mild encounters.

The inner-disk spiral arms of IC 2163 may be an earlier form of the
very bright inner-disk spiral arms of NGC 5394, discussed in Section
\ref{sect:intro}. The model in Kaufman et al.  (1999) finds that the
latter developed from an ocular structure present at a slightly earlier
time. NGC 5394 has a central starburst and its inner-disk spiral arms
are unusual in that two of the three bright arms show no evidence for
star formation.

\section{Peculiar Emission Structures}
\label{sect:pec}

We examined the {\it HST} image in detail for all peculiar emission
features that might be conical or jet-like. Several interesting candidates
were found. Most have a size in the range from 100 to 1000 pc.

\subsection{Features on 100 to 1000 pc Scales} 

A collection of peculiar emission features on 100 to 1000 pc scales is
found in Figure \ref{fig:jets} and a finding chart is given in Figure
\ref{fig:find}. North is up for all the images in Figure \ref{fig:jets}
and the physical scale is on the right, assuming a galaxy distance of
35 Mpc.  No feature appears in only one filter, which would make it an
image flaw.

There are several linear or arc-shaped structures of blue knotty emission,
such as c, h, s, t, and o. They could be composed of young stars except
for their peculiar linear shapes. Feature f is similar to these but
brighter; u is a curved line of star formation. Features a, g and p are
other multiple-point structures, apparently made from young stars.

Jet-like features are seen in several places: e contains two jet-like
structures that point toward each other; q has two jet-like features
with a star-like object in the middle; b has a small jet-like object
pointing away from a bright diffuse patch and other linear structures
perpendicular to this; g (mentioned above) also contains a streak of
faint emission between the central pointy structure and a star-like
object in the north; n contains three faint linear emission streaks in
what seems to be a region of star formation.

V-shaped features that could be conical emission regions include:
i, the strong radio continuum source in the far west of NGC 2207,
containing intense star formation, a dense dust cloud and a possible
conical emission region in the north (see Sect. \ref{sect:rcsource}); l,
a sideways V-shape, possibly a conical dust and reflection feature in the
NGC 2207 spiral arm with blue star formation knots in it; and possibly
r, a peculiar dark feature just to the upper right of the center, with
a bright rim around it. Feature j contains two nearly parallel streaks
running vertically in the figure, one streak is bright and the other is
dark, giving it a three-dimensional quality.

There are two dust shells: k, a dust shell with three star formation knots
in it, most likely in the NGC 2207 spiral because of the blue color of
the star formation; and m, a dust shell with a red star in the middle
(possibly a foreground star, see Paper V).

To the northwest of the dust shell k, there is a bright kpc-long arc
(feature d) that looks like reflection off the inner edge of the dust
lane that is inside the inner-disk arm.

The reddish features (k, d, l, m, n) juxtaposed on top of IC 2163 could
be inside IC 2163 or in the foreground spiral arms of NGC 2207. The
other features are mostly white or blue in color, and, except for a and
possibly r, are positioned to be likely members of NGC 2207.

Individual jet-like objects are typically
$1^{\prime\prime}-2^{\prime\prime}$ or 150-300 pc long; the two streaks in
feature e span 1000 pc.  Analogous objects are not known in our Galaxy;
much smaller versions of jets could be those associated with SS 443
(Margon 1984), the Crab Nebula jet (van den Bergh 1970; Gull \& Fesen
1983), or Cas A (Fesen \& Gunderson 1996). The jet-like regions in NGC
2207 might also be radio sources that are too weak and small to have
been seen in our VLA survey.  The origin of these features is not clear.
Some may be related to mass transfer events in which gas from IC 2163
recently impacted the disk of NGC 2207, some may be optical illusions,
and some may be normal galactic features not previously recognized in
other galaxies.

\subsection{Larger-Scale Linear Features} 

The top panel of Figure \ref{fig:jets} shows four unusual elongated
features that lie on approximately the same line. Each cuts across a
spiral arm at a large angle. The most prominent is the straight ridge
of star formation along the innermost northern spiral arm of NGC 2207
(near the middle of this panel at $\alpha_{1950} = 6^h 14^m 15.7^s$,
$\delta_{1950} = -21^\circ 21^\prime 5^{\prime\prime}$).  This ridge
alone is not so peculiar because it has a dust lane on the inner edge and
normal--looking young stars, but it is unusually straight for a density
wave feature, and it does not follow the curvature of the rest of the
arm. Just to the west of this bright feature (and along the same line)
is a faint, thinner emission streak composed of several diffuse spots
with no obvious dust lane. It cuts across the same spiral arm. In the
opposite direction, in the next outer arm of NGC 2207 to the east of
the brightest linear streak, there is another linear feature composed
of dust in the center of two streaks of star formation; this is also not
aligned with its local arm. The fourth feature is in the middle western
arm of NGC 2207, on the right side of the top panel at $\alpha_{1950} =
6^h 14^m 10.7^s$, $\delta_{1950} = -21^\circ 20^\prime 47^{\prime\prime}$
(enlarged in panel u). The linear part crossing the spiral arm is made
of dust and star formation, but just to the west of the arm, in the
interarm region, is a large, diffuse ``bubble-like'' object. Several
blue emission spots and a dust filament curve upward from the end of
the bubble, and a faint, sideways V-shaped object that looks like a
bow shock is further to the west, along the same line (cf. feature u in
Fig. \ref{fig:jets}). The whole line of 4 features points back to within
1$^{\prime\prime}$ of the nucleus of IC 2163. This alignment resembles
a jet, but there is no prominent radio continuum source in the center
of IC 2163 nor along the line. Paper III found a strong radio continuum
ridge $\sim 15''$ north of the line connecting these four linear features,
and aligned at an angle of $30^\circ$ relative to it.

The original WFPC2 chip seam on the mosaic runs underneath and almost
parallel to this linear structure, beginning at the bright emission knot
in the arm to the east of the main ridge and ending at the edge of the
image on the west. However the individual features are visible in the
separate fields before the mosaic was made, so these features are not
artifacts of the seam.

Detailed numerical models of both galaxies in the interaction (Struck
et al. 2000) occasionally find linear features from debris trails with
material pulled out of IC 2163 and brushing the backside of NGC 2207.
The lifetimes of any linear features composed of stars and dust are
constrained by galactic rotation to be fairly short, less than several
million years.

\section{H~I Emission and Star Formation} 
\label{sect:hi}

Maps of the H~I emission from each galaxy are shown superposed on the
{\it HST} images in Figures \ref{fig:himap2163} and
\ref{fig:himap2207}. Two peculiarities of the H~I emission are its
large velocity dispersion ($\sim50$ km s$^{-1}$) and large cloud masses
($\sim10^8$ M$_\odot$; Paper III), which are characteristic of interacting
systems (Paper II; Kaufman et al. 1997; 1999; Irwin 1994; Hibbard 1995).

The largest clouds in these galaxies are not clearly associated with
star formation. Figure 14 in Paper III was a finding chart for $10^8$
M$_\odot$ clouds that were also outlined on the contour diagrams in
Figures 8 and 16 of that paper. The same contours are shown here in
Figures \ref{fig:himap2163} and \ref{fig:himap2207}, but without the
outlines for giant clouds. Here the clouds in NGC 2207 are identified
by the notation N1, N2, and so on.

In IC 2163, there are five giant concentrations of H~I, two in the eastern
tidal arm, one in the streaming region of the eastern tidal arm to the
north, one in the western tidal arm near the center of NGC 2207, and
one in the northern eyelid region (cf. Fig. \ref{fig:himap2163}). None
are associated with specific star formation regions, but the one on the
northern eyelid has star formation all along its inside border. The one
in the western tidal arm is heavily obscured by NGC 2207. There is a
less massive H~I cloud in the southern eyelid region which is centered
on a region with bright patches of star formation.

In NGC 2207 there is a giant H~I cloud in the far northwest (N1)
with only faint emission in the B band and three more in the western
spiral arm below this (N2, N3, N4) with bright star formation only
in the lowest one. The other two giant clouds in NGC 2207 (N5, N6)
are in the east in Figure \ref{fig:himap2207}, in the region that is
foreground to IC 2163. One (N5) is at the end of the outer eastern arm
of NGC 2207 and is associated with a dust feature and faint blue cluster
(cf. Fig. \ref{fig:mainimage}).  The other (N6) is in a region crossed
by dust lanes between the tidal bridge of IC 2163 and the middle arm
on the eastern side of NGC 2207. The cloud is not centered on a bright
star-forming region. It coincides with part of a 10 kpc long, radio
continuum ridge (see Paper III).

The bright star-forming regions are mostly associated with smaller H~I
clouds, having masses of $10^7$ M$_\odot$ or less. At this level of H~I
emission, many star-forming regions have some H~I nearby, as is often the
case in non-interacting galaxies. For example, in NGC 2207, the small
patches of star formation in the southern arm are all associated with
these less massive H~I clouds, as is the small patch south of feature i
(Fig. \ref{fig:jets}) in the western arm.  The middle arm on the western
side of NGC 2207 winds around to the north and then crosses in front of
the central part of IC 2163.  In the north, this arm has several bright
regions of star formation along the H~I ridge, some of which coincide
with H~I clumps. On the eastern side, the arm, seen as the dust streamers
discussed in Section 4.3, also coincides with the H~I ridge.

The giant star-forming region (feature i) that is associated with the
intense radio continuum source on the outer western arm has no H~I cloud,
although there are smaller clouds to the south and west of it (N3, N2).
The dense dust in this region suggests the hydrogen is present in
molecular form.

There is no direct optical evidence for the large H~I velocity dispersions
that are present in these galaxies, which is typically 30--50 km s$^{-1}$
(Paper II).  Such motions would be supersonic for H~I.  Large random
motions also imply that the gas disks are thicker than normal, by
perhaps a factor of $\sim5$.  The models in Paper IV predict a warp in
NGC 2207 that is pointed away from us on the IC 2163-side and toward
us on the western side. Paper III found that on the western side of NGC
2207, the optical radial surface brightness profile has a plateau from
$40^{\prime\prime}-90^{\prime\prime}$, which corresponds to the broad
ring of H~I emission that peaks at $70^{\prime\prime}$.  The increase
in the line-of-sight thickness produced by the warp may be responsible
for the prominence of the H~I ring on the eastern and western sides and
for the unusual behavior of the radial surface brightness profile in
the optical on the western side.  Since most of the massive clouds in
NGC 2207 are in these directions, their surface mass densities would be
slightly smaller than computed by neglecting the warp.

The high H~I velocity dispersion could affect the optical observations
in other ways too. The dispersion could broaden the gas response to
the spiral arms, for example, or remove the tendency for the gas to
shock once in a thin dust lane. Instead, it could be shocking several
times in the parallel dust lanes that are observed in front of IC 2163
(see Sect.  \ref{sect:foreground}).  The H~I line profiles are typically
not Gaussian, so they could be composed of a small number of streams
with different velocities.  In that case, the high velocity dispersion
could result from variations in the systematic motions within the H~I
synthesized beam. These systematic motions could lead to the multiple
shocks seen as dust lanes. Because the H~I observations have a large beam
($13.5^{\prime\prime}\times12^{\prime\prime}$), any streaming motions
on a kpc scale would be unresolved and appear only as turbulence.

Strong turbulence or high speed streaming motions could also make the
arms in NGC 2207 thicker than those in other spiral galaxies, such as
M100 or M81. The NGC 2207 arms are similar to those in Luminosity Class
III galaxies, which are usually much smaller than NGC 2207 (Iye \&
Kodaria 1976). This makes sense if the relative thickness of the arms
scales with the ratio of the velocity dispersion to the orbit speed.

The origin of the high HI velocity dispersion is not known.  Other
strongly interacting galaxies have these motions too, so the turbulence
could come from internal adjustments to tidal forces. We show in Paper
V that the regions of highest dispersion contain superstar clusters,
so there could be a connection with star formation too.

\section{Conclusions}

The galaxies IC 2163 and NGC 2207 are involved in a near encounter that
has compressed IC 2163 in the plane and warped NGC 2207 out of the plane.
Optical observations with {\it HST} show many interesting and peculiar
features. Two types of extinction patterns were discussed in Section
\ref{sect:dust}: long parallel filaments in the tidal tail of IC 2163,
and clumpy filaments in the spiral arms of NGC 2207 that are seen in
projection against IC 2163.

The tidal-tail filaments seem to be normal flocculent spiral arms
that were in the disk of IC 2163 before the encounter and then got
stretched out into the tidal tail after the encounter by the overall
disk deformation. A numerical simulation reproduced these filaments
well. The extreme youth of the interaction ($\sim40$ My) explains why the
spiral arms were not individually affected much, except for the overall
distortion that followed the tidal flow. Many of the filaments have star
formation that trails systematically behind them, suggesting a large-scale
acceleration of the gas relative to the stars that form inside it.

The foreground spiral arm in NGC 2207 shows multiple parallel filaments
as well, but probably for a different reason. There could be multiple
shocks in the density wave, or independent filaments that came in from
the interarm region. Such structure could be normal for galaxies but
not commonly observed because of the rarity of background lighting. In
any case, star formation occurs in the clumps of these filaments in
a way that resembles local star formation in small filamentary dark
clouds. The process of star formation in the density waves of NGC 2207 is
therefore one in which filaments, either made by the wave or pre-existing,
collapse gravitationally into globules which then continue to collapse
into clusters and individual stars. The star formation does not look
like it is occurring at the interfaces between colliding clouds or
colliding filaments.

Another interesting region is a dense dark cloud in a region of star
formation on the outer western arm of NGC 2207.  This is the site
of the most luminous H$\alpha$ source in the galaxies, and a large,
nonthermal radio continuum source 1500 times more luminous that Cas A.
There is a massive cluster in the center, a conical feature like an
outflow to the north, and other clusters in the southeast that could
have been triggered by energetic explosions.

The spirals inside the oval of IC 2163 were discussed in Section
\ref{sect:nucsp}. The dynamical time of this inner region is much
shorter than the interaction time, so these spirals could be part of
the response. We suggested, on the basis of the rotation curve, that the
central spirals could extend from the outer ILR to the inner ILR for a
pattern speed that places corotation at the companion.  In that case,
the whole oval may be viewed as a bar-like stellar+gaseous flow pattern
in the transient $\cos(2\theta)$ potential of the tidal field. The
inner spirals are then a normal resonance response for such a bar. This
circumstance highlights the interesting possibility that transient tidal
forces may trigger nuclear gas inflow via bar-like hydrodynamics even
when there is no bar. It also suggests that post-encounter starburst
systems without obvious bars today may have had their major accretion
events close to perigalacticon when the orbits were highly distorted.

Numerous peculiar emission features were found after a close examination
of the {\it HST} image. Many are linear with either smooth or clumpy
internal structures, and some have star-like objects at one or both
ends. They are typically several hundred parsecs long, although smaller
features would be difficult to see. Some could be jets or conical
outflows, but they are much larger than protostellar jets found in the
Milky Way. They may be coincidental alignments of stars or clusters, but
the smoothest ones do not look like this.  A long line of four co-linear
features that are at odd angles to their local spiral arms was also noted.

The H~I is weakly associated with star formation in these galaxies,
but the largest H~I clouds, with masses of around $10^8$ M$_\odot$,
are not generally producing rich clusters. The H~I velocity dispersion
is $5\times$ higher than normal, which may explain why the spiral arms
in NGC 2207 look thick and feathery.

The high-resolution images obtained by the {\it Hubble Space Telescope}
have presented us with an unprecedented opportunity to study the
morphology of a pair of interacting galaxies in the early stages of
interaction. It has shown us detailed features not previously observed
elsewhere, and given us important insights into the early stages of
galaxy interactions.

Acknowledgements: 

This work was supported by NASA through grant number GO-06483-95A from the
Space Telescope Science Institute, which is operated by the Association
of Universities for Research in Astronomy, Inc., under NASA contract
NAS5-26555.  Support for the Hubble Heritage Team's contribution also
came from STScI.

\newpage
\begin{figure}
\figcaption{Interacting galaxies IC 2163 (East) and NGC 2207
(West), from a mosaic of three {\it HST} pointings. North is to the left in this print.
Resolution degraded for astro-ph.}
\label{fig:mainimage}
\end{figure}

\begin{figure}
\figcaption{Numerical models of the parallel filaments in the tidal
arm of IC 2163 (a) at the beginning of the interaction, (b) 
at the time
best representing the current epoch, (c) velocity 
vectors at the current epoch, (d) a projected view of the 
density for comparison with the image of IC 2163 in the sky, rendered as
an 160x160 array of pixels. In a-c, color represents initial position
in the disk; in d, color represents density, going from blue to red
with increasing density.}
\label{fig:model}
\end{figure}

\begin{figure}
\figcaption{
H~I column density contours of IC 2163 (from Paper III) with
a resolution of $13.5^{\prime\prime} \times 
12^{\prime\prime}$ (FWHM) overlaid on the {\it HST} B band
image in grayscale. Contour levels of
the line-of-sight H~I column density are 5, 10, 15, 20, and 25 M$_\odot$ 
pc$^{-2}$.
On the eastern side of IC 2163, triangles on the tidal tail mark the ridge
of the high-velocity H~I streaming component, pentagons mark the ridge of
the low-velocity H~I component, and plus signs mark the dividing line between
the high and low velocity H~I ridges. On the western
side of IC 2163, there appears to be a faint blue arm coincident
with the H~I tidal bridge.  }
\label{fig:himap2163}
\end{figure}

\begin{figure}
\figcaption{An enlargement of the region containing the
nonthermal radio continuum source on the western arm of NGC 2207
(cf. panel i in Figs. 6 and 7). The contours are radio continuum emission at
$\lambda$ = 20 cm with a resolution of $10^{\prime\prime}
\times 6.5^{\prime\prime}$ (FWHM).
They are overlaid
on the {\it HST} B band image in grayscale. The contour interval is 10 K, which 
is 3 times
the rms noise, and the contour levels are at 1, 2, 3, 4, 5, and 6 times the
contour interval. The plus sign marks the location of
the radio continuum maximum in the higher resolution observations by
Vila et al. (1990).}
\label{fig:rcsources}
\end{figure}

\begin{figure}
\figcaption{Arm/interarm contrast as a function of radius for the
inner spirals in IC 2163. The lack of an increase with radius
suggests a location inside the ILR.}
\label{fig:ai}
\end{figure}

\begin{figure}
\figcaption{Montage of peculiar features showing linear or smooth
emission structures. Physical scale on the right assumes a distance
of 35 Mpc. }
\label{fig:jets}
\end{figure}

\begin{figure}
\figcaption{Finding chart for the peculiar features shown in Figure 6.}
\label{fig:find}
\end{figure}

\begin{figure}
\figcaption{H~I column density contours of NGC 2207 (from Paper III) with
a resolution of $13.5^{\prime\prime}\times 12^{\prime\prime}$ (FWHM) overlaid on the {\it HST} B band
image in grayscale. Contour levels of the line-of-sight H~I column density are
10, 15, 20, 25, 30, and 35 M$_\odot$ 
pc$^{-2}$. The dust streamers in the backlit
spiral arm of NGC 2207 are discussed in Section 4.3 lie along an H~I ridge.  }
\label{fig:himap2207}
\end{figure}

\begin{table}                                                                   
\caption{Basic Data on IC~2163/NGC~2207}                                        
\begin{center}                                                                  
\begin{tabular}{lcc}                                                            
Characteristic        &\multispan{2}\hfill Galaxy\hfill\\                       
&IC~2163       &NGC~2207\\                                                      
\tableline                                                                      
Morphological type      &SB(rs)c pec   &SAB(rs)bc pec\\                         
Right Ascension (1950)      &$6^h14^m20.0^s$  &$6^h14^m14.4^s$\\                
Declination (1950)    &-21$^{\circ}21^{\prime}24^{\prime\prime}$
&-21$^{\circ}21^{\prime}14^{\prime\prime}$\\  
Right Ascension (2000)      &$6^h16^m27.7^s$  &$6^h16^m22.1^s$\\                
Declination (2000)    &-21$^{\circ}22^{\prime}1^{\prime\prime}$
&-21$^{\circ}22^{\prime}21^{\prime\prime}$\\  
Isophotal Major Radius (R$_{25}$)     &$1.51^{\prime}$ &$2.13^{\prime}$\\                      
Axis ratio                         &2.45      &1.55  \\                         
$m_B$                              &12.55     &11.59 \\                         
Corrected B magnitude, $B_T^0$    &11.26     &10.90 \\                          
Distance                    &\multispan{2}\hfill 35 Mpc\hfill \\                
Velocity (km s$^{-1})$&$2765\pm20$&$2745\pm15$\\
\end{tabular}                                                                   
\end{center}                                                                    
\end{table}

\end{document}